\documentclass[12pt]{article}

\usepackage{epsf}
\usepackage{wick}
\usepackage{amsmath}
\allowdisplaybreaks

\begin{document}

\baselineskip=18.6pt plus 0.2pt minus 0.1pt

%%%%%%%%%%%%%%%%%%%%%%%%%%%%%%%%%%%%%%%%%%%%%%%%%%%%%%%%%%%%%%%%
\makeatletter
\@addtoreset{equation}{section}
\renewcommand{\theequation}{\thesection.\arabic{equation}}
\renewcommand{\thefootnote}{\fnsymbol{footnote}}
%%%%%%%%%%%%%%% Private Macros %%%%%%%%%%%%%%%%%%%%%%%%%%%%%%%%%%
\newcommand{\p}{\partial}
\newcommand{\bra}[1]{\left\langle #1\right\vert}
\newcommand{\ket}[1]{\left\vert #1\right\rangle}
\newcommand{\nn}{\nonumber}
\newcommand{\CO}[2]{[#1,#2]}
\newcommand{\PB}[2]{\{#1,#2\}}
\newcommand{\unit}{\hbox to 3.8pt{\hskip1.3pt \vrule height 7.4pt
    width .4pt \hskip.7pt \vrule height 7.85pt width .4pt \kern-2.4pt
    \hrulefill \kern-3pt \raise 3.7pt\hbox{\char'40}}}
\newcommand{\xh}{\widehat{x}}
\newcommand{\tr}{\mathop{\rm Tr}}
\newcommand{\hx}{\widehat{x}}
\newcommand{\QQ}{\theta^2}
\newcommand{\vv}{\mbox{\boldmath$v$}}
\newcommand{\tx}{(\theta\hx)}
\newcommand{\hA}{\widehat{A}}
\newcommand{\hd}{\widehat{\delta}}
\newcommand{\hl}{\widehat{\lambda}}
\newcommand{\D}{\cal D}
\newcommand{\hy}{\widehat{y}}
\newcommand{\ty}{(\theta\hy)}
\newcommand{\hQ}{\widehat{\theta}}
\newcommand{\hF}{\widehat{F}}
\newcommand{\hD}{\widehat{D}}
\newcommand{\hPhi}{\widehat{\Phi}}
\newcommand{\R}{\mathop{\mbox{Re}}}
\newcommand{\I}{\mathop{\mbox{Im}}}
\newcommand{\pA}{\p{}A}
\newcommand{\ppA}{\p\p{}A}
\newcommand{\piA}{\p_i{}A}
\newcommand{\pF}{\p{}F}
\newcommand{\wt}{\widetilde}
\newcommand{\wh}{\widehat}
\newcommand{\calO}{{\cal O}}
\newcommand{\calD}{{\cal D}}
\newcommand{\calF}{{\cal F}}
\newcommand{\ev}{\Lambda}
\def\contr#1#2#3#4{\vbox{\ialign{##\crcr
          \hskip #2pt\vrule depth #4pt
          \hrulefill\vrule depth #4pt\hskip #3pt
          \crcr\noalign{\kern-1pt\vskip0.125cm\nointerlineskip}
          $\hfil\displaystyle{#1}\hfil$\crcr}}}
\newcommand{\scF}{\contr{F}{2}{2}{4}}

\makeatother
%%%%%%%%%%%%%%% End of Private Macros %%%%%%%%%%%%%%%%%%%%%%%%%%

\begin{titlepage}
\title{
\hfill\parbox{4cm}
{\normalsize KUNS-1663\\{\tt hep-th/0005101}}\\
\vspace{1cm}
Noncommutative Monopole at the Second Order in $\theta$
}
\author{
Seiya {\sc Goto}\thanks{{\tt goto@gauge.scphys.kyoto-u.ac.jp}}
{} and
Hiroyuki {\sc Hata}\thanks{{\tt hata@gauge.scphys.kyoto-u.ac.jp}}
\\[7pt]
{\it Department of Physics, Kyoto University, Kyoto 606-8502, Japan}
}
\date{\normalsize May, 2000}
\maketitle
\thispagestyle{empty}

\begin{abstract}
\normalsize\noindent
We study the noncommutative $U(2)$ monopole solution at the second
order in the noncommutativity parameter $\theta^{ij}$.
We solve the BPS equation in noncommutative super Yang-Mills
theory to $\calO(\theta^2)$, transform the solution to the
commutative description by the Seiberg-Witten (SW) map, and evaluate
the eigenvalues of the scalar field.
We find that, by tuning the free parameters in the SW map,
we can make the scalar eigenvalues precisely reproduce the
configuration of a tilted D-string suspended between two parallel
D3-branes. This gives an example of how the ambiguities inevitable in
the higher order SW map are fixed by physical requirements.

\end{abstract}

\end{titlepage}

\section{Introduction}

An explicit relation between the noncommutative fields and
the commutative ones has been presented in \cite{SW}, called
``the Seiberg-Witten (SW) map.''
The noncommutative Dirac-Born-Infeld (DBI) theory and the ordinary one
appear as low-energy effective theories of D-brane in a constant
NSNS $B$-field.
They differ by the choice of the regularization for the worldsheet
theory;
the Pauli-Villars regularization for the commutative description
and the point-splitting regularization for the noncommutative one.
This means that these two descriptions are connected by some field
redefinition and this is the SW map.

The relation between the commutative and noncommutative DBI theories
has been examined in various aspects.
Now in particular, let us concentrate on the BPS solutions and compare
them in both the descriptions.
The reason is that the BPS solutions are considered as powerful
tools beyond the perturbative understanding.
Noncommutative BPS monopoles describe, by the brane interpretation of
\cite{CM}, the configurations of tilted D-strings ending on parallel
D3-branes in a constant NSNS $B$-field \cite{HH} and have been
investigated in various papers \cite{HHM,Bak,HM,HaHi,Mori}.
In \cite{HHM}, noncommutative $U(2)$ monopole was considered at the
first order in the noncommutativity parameter $\theta^{ij}$.
The analysis using the noncommutative eigenvalue equation for the
scalar field successfully reproduced the tilted D-string picture.
In \cite{HM}, the similar analysis was carried out for the string
junction and the anticipated result was obtained.
Study of the noncommutative monopoles using the SW map was carried
out in \cite{HaHi,Mori} (see also \cite{Mat}).
There, the noncommutative BPS solutions were transformed into the
commutative description via the SW map, and then the brane
interpretation was done for the eigenvalues of the mapped scalar
field to give the expected tilted D-string picture.

The purpose of this paper is to extend the analysis of the
noncommutative $U(2)$ monopole using the SW map to second order in
$\theta$.
The motivating fact is as follows: the SW map possesses
some ambiguities in higher orders in $\theta$ \cite{AK}.
This map is derived from the requirement of the gauge equivalence of
the two descriptions. Since this is a very weak requirement,
arbitrary parameters appear in the map. There are two types of
ambiguities in it.
One is in the form of the gauge transformation and has no physical
effect.
However, the other type of ambiguity consists of gauge covariant
quantities and can cause physical differences.

We apply the SW map to the noncommutative monopole
solution at the second order in $\theta$ and examine the effects of
the ambiguities. Concretely, we compare the eigenvalues of the scalar
field obtained by the SW map with that in the commutative Yang-Mills
theory in a background magnetic field.
Note that the ambiguities in the SW map can change the scalar
eigenvalues (which are gauge invariant quantities) and hence change
their brane interpretation. It is found that
we can make these two eigenvalues coincide with each other by tuning
the free parameters in the SW map.
This gives an example of how the ambiguities in the SW map are fixed
in concrete physical situations.

The rest of this paper is organized as follows.
In section 2, we solve the noncommutative version of the BPS
equation to second order in $\theta$. In section 3, we apply
the SW map to the solution and evaluate the eigenvalues of the scalar
field in the commutative description.
In section 4, we compare the scalar eigenvalues of section 3 with
those in the commutative Yang-Mills theory in a constant magnetic
field, and examine the effect of the ambiguities in the SW map.
In section 4, we summarize the paper and give some discussions.
The SW map to second order in the change of $\theta$ is presented
in Appendix A.

\section{Noncommutative BPS monopole solution at $\theta^2$}

We shall consider the ${\cal N}=4$ $U(2)$ noncommutative super
Yang-Mills theory in $1+3$ dimensions with the metric
$G_{\mu\nu}={\rm diag}(-1,1,1,1)$, and construct the BPS monopole
solution to second order in the noncommutativity parameter.
The BPS equation reads
\begin{equation}
\hD_i\hPhi+\frac{1}{2}\epsilon_{ijk}\hF_{jk}=0,
\label{BPSeq}
\end{equation}
where the quantities with a hat denote those in the noncommutative
description. In particular, we have
\begin{align}
\hF_{ij}&\equiv \p_i{}\hA_j-\p_j{}\hA_i
-i \hA_i{}*{}\hA_j{}+i\hA_j * \hA_i ,\\
\hD_i \hPhi&\equiv \p_i \hPhi - i\hA_i * \hPhi + i \hPhi * \hA_i ,
\end{align}
where the $*$ product is defined by
\begin{align}
(f*g)(x)&\equiv
f(x)\exp\left(\frac{i}{2}\theta^{ij}
\overleftarrow{\p_i}\overrightarrow{\p_j}\right)g(x)\nn\\
&=f(x)g(x)+ \frac{i}{2}\theta^{ij}\p_if(x)\,\p_jg(x)
-\frac{1}{8}\theta^{ij}\theta^{kl}\p_i\p_kf(x)\,\p_j\p_lg(x)
+\calO(\theta^3).
\label{star}
\end{align}

In order to solve the BPS equation (\ref{BPSeq}),
we expand the fields in powers of $\theta$:
\begin{align}
\hPhi &\equiv
\left(
\hPhi^{a(0)}+\hPhi^{a(1)}+\hPhi^{a(2)}
\right)\frac{1}{2}\sigma_a
+
\left(
\hPhi^{0(1)}+\hPhi^{0(2)}
\right)\frac{1}{2}{\unit},\nn\\
\hA_i &\equiv
\left(
\hA_i^{a(0)}+\hA_i^{a(1)}+\hA_i^{a(2)}
\right)\frac{1}{2}\sigma_a
+
\left(
\hA_i^{0(1)}+\hA_i^{0(2)}
\right)\frac{1}{2}{\unit},
\label{expand}
\end{align}
where the superscript $(n)$ denotes the order of $\theta$.
As the solution at $\calO(\theta^0)$, we adopt the
BPS monopole \cite{PS,Bogo} with vanishing $U(1)$ components
$\hPhi^{0(0)}=\hA^{0(0)}_i=0$:
\begin{equation}
\hPhi^{a(0)} = \frac{\xh_a}{r}H(\xi),\quad
\hA_i^{a(0)} = \epsilon_{aij}\frac{\xh_j}{r}
\bigl(1-K(\xi)\bigr),
\label{t0 sol}
\end{equation}
with $\hx_i\equiv{}x_i/r$ and $\xi\equiv{}Cr$
($C$ is the parameter characterizing the mass of the monopole).
The functions $H$ and $K$ are defined by
\begin{equation}
H(\xi)=\frac{\xi}{\tanh\xi}-1,\quad K(\xi)=\frac{\xi}{\sinh\xi}.
\end{equation}
These two functions behave asymptotically as
\begin{equation}
H(\xi)=\xi-1+{\cal O}(e^{-\xi}),\quad
K(\xi)=0+{\cal O}(e^{-\xi}).
\label{asympt}
\end{equation}
Note that the solution (\ref{t0 sol}) has an invariance
under the rotation by the diagonal subgroup $SO(3)$ of
$SO(3)_{\mathrm{space}}\otimes{}SO(3)_{\mathrm{gauge}}$.

The solution at $\calO(\theta)$ was constructed in \cite{HHM,Bak}
and it is given by
\begin{alignat}{2}
\hA^{a(1)}_i&=0,&\qquad
\hA^{0(1)}_i &= \theta^{ij}\hx_j \frac{1}{4r^3}(1-K)(1-K+2H),\nn\\
\hPhi^{a(1)}&=0,&\qquad
\hPhi^{0(1)} &= 0.
\label{t1 sol}
\end{alignat}
This solution is invariant under the generalized rotation,
namely, the simultaneous rotation of the diagonal $SO(3)$ and
the indices of the noncommutativity parameter $\theta_{ij}$.
Note that the noncommutativity has no influence on the scalar solution
at $\calO(\theta)$.

Now let us consider the components at the second order in $\theta$ in
the expansion (\ref{expand}).
The $\calO(\theta^2)$ part of the BPS equation reads
\begin{align}
&\p_i\hPhi^{0(2)}+\epsilon_{ijk}\p_j \hA_k^{0(2)}=0,
\label{U1}
\\[1.5ex]
&\p_i\hPhi^{a(2)}
+\epsilon_{ijk}\p_j\hA_k^{a(2)}
+\epsilon_{abc}\left(
\hA_i^{b(0)}\hPhi^{c(2)}-\hPhi^{b(0)}\hA_i^{c(2)}
+\epsilon_{ijk}\hA_j^{b(0)}\hA_k^{c(2)}\right)
\nn\\
&\qquad
=
\frac12\theta^{kl}\p_k\hPhi^{a(0)}\p_l\hA_i^{0(1)}
-\frac12\epsilon_{ijk}\theta^{lm}
\p_l\hA_j^{a(0)} \p_m\hA_k^{0(1)}
\nn\\
&\qquad\quad
+\frac18\epsilon_{abc}\theta^{lm}\theta^{pq}\left(
\p_l\p_p\hA_i^{b(0)} \p_m\p_q\hPhi^{c(0)}
+\frac12\epsilon_{ijk}
\p_l\p_p\hA_j^{b(0)} \p_m\p_q\hA_k^{c(0)}
\right) ,
\label{SU2}
\end{align}
where the first equation (\ref{U1}) is the $U(1)$ part of
(\ref{BPSeq}), while the second equation (\ref{SU2}) is the $SU(2)$
part. The $U(1)$ part has no regular solutions and we shall
concentrate on the $SU(2)$ part (\ref{SU2}).

In order to solve eq.\ (\ref{SU2}),
we adopt the generalized rotational invariance used in the
construction of the ${\cal O}(\theta)$ part (\ref{t1 sol}), and expand
$\hPhi^{a(2)}$ and $\hA_i^{a(2)}$ as
\begin{align}
\hPhi^{a(2)}&=\frac{1}{r^5}\Bigl[
\phi_1(\xi)\tx\theta_a +\phi_2(\xi)\QQ\hat{x}_a
+ \phi_3(\xi)\tx^2\hat{x}_a
\Bigr],\nn\\[1.5ex]
\hA_i^{a(2)}&=\frac{1}{r^5}\Bigl[
a_1(\xi)\QQ\epsilon_{aij}\hx_{j}+a_2(\xi)\tx\epsilon_{aij}\theta_j
+ a_3(\xi)\epsilon_{ajk}\theta_{i}\theta_{j}\hx_{k}
\nn\\
&\qquad\quad
+ a_4(\xi)\tx^2\epsilon_{aij}\hx_{j}
+ a_5(\xi)\tx\epsilon_{ajk}\hx_i\theta_j\hx_k
\Bigr] ,\label{t2 sol}
\end{align}
where we have used
$\theta_i\equiv(1/2)\epsilon_{ijk}\theta^{jk}$,
$\QQ\equiv\theta_i\theta_i$ and $\tx\equiv\theta_i\hx_i$.
One can check that this is the most general expression
satisfying the generalized rotational invariance, by using the
identities,
\begin{equation}
\epsilon_{aij}x^2
=\left(
\epsilon_{kij}x_a+\epsilon_{akj}x_i+\epsilon_{aik}x_j
\right)x_k,
\end{equation}
and the same one with $x_i$ replaced by $\theta_i$.
Putting the expansion (\ref{t2 sol}) into (\ref{SU2}),
we obtain the following linear differential equations with
inhomogeneous terms for the unknown functions $\phi_k(\xi)$ and
$a_k(\xi)$:
\begin{align}
&\calD(-a_1-a_3)+\phi_2 +4 a_1 - a_2 + 5 a_3 + a_5
-(1-K)\phi_2 - H a_1 = I_1 ,
\nn\\
&\calD(\phi_2 + a_1 + a_3)- 6 \phi_2- 6 a_1 - 6 a_3 - a_5
+(1-K)(\phi_2 + 2 a_1 + a_3) + H a_1 = I_2 ,
\nn\\
&\calD a_3 + \phi_1 + a_2 - 5 a_3 -a_5 - H a_3 =I_3 ,
\nn\\
&\calD(a_2 - a_3) + 2 \phi_3 -6 a_2 + 6 a_3 + 2 a_5
+(1-K)(\phi_1 + a_2-a_3-a_5) + H a_3 = I_4 ,
\nn\\
&\calD(\phi_1 - a_3)- 6 \phi_1 + 6 a_3 + 2 a_4 + a_5
+(1-K) a_2 + H(a_2 - a_5) =I_5 ,
\nn\\
&\calD(-a_2+a_3 -a_4)+ \phi_3 + 6 a_2 - 6 a_3 + 4 a_4 - 2 a_5
-(1-K)(\phi_1+\phi_3)-H(a_2 + a_4) = I_6 ,
\nn\\
&\calD(\phi_3 + a_4) - 8 \phi_3 - 8 a_4
+(1-K)(\phi_3 + 2 a_4 + a_5)+H(a_4 + a_5) = I_7 ,
\label{diffeq}
\end{align}
where the differential operator ${\cal D}\equiv\xi(d/d\xi)$ has been
introduced.
The seven equations (\ref{diffeq}) correspond to seven
independent structures of eq.\ (\ref{SU2});
$\QQ\delta_{ai}$, $\QQ\hx_a\hx_i$, $\theta_a\theta_i$,
$\tx\hx_a\theta_i$, $\tx\theta_a\hx_i$, $\tx^2\delta_{ai}$ and
$\tx^2\hx_a\hx_i$, respectively.
The functions $I_k$ ($k=1,\cdots,7$) are the inhomogeneous terms
which are polynomials of $H$ and $K$:
\begin{align}
I_1(\xi)
&=
-\frac{13}{8}+\frac{3H}{8}-\frac{H^2}{2}
+\frac{9K}{4}+\frac{5HK}{4}+\frac{3H^2K}{8}
+\frac{H^3K}{8}+\frac{K^2}{8}-\frac{5HK^2}{8}\nn\\
&\quad
-\frac{H^2K^2}{4}-\frac{5K^3}{8}-HK^3-\frac{3H^2K^3}{8}
-\frac{K^5}{8},\nn\\
I_2(\xi)
&=
\frac{3}{8}+\frac{7H}{8}+\frac{H^2}{2}-\frac{9K}{8}
-\frac{3HK}{2}-\frac{3H^2K}{8}-\frac{H^3K}{8}+K^2
+\frac{HK^2}{8}\nn\\
&\quad
-\frac{H^2K^2}{2}-\frac{H^3K^2}{4}
+\frac{3HK^3}{4}+\frac{3H^2K^3}{8}-\frac{3K^4}{8}
-\frac{HK^4}{4}+\frac{K^5}{8},\nn\\
I_3(\xi)
&=
\frac{5}{8}+\frac{9H}{8}+\frac{H^2}{2}-\frac{17K}{8}-3HK
-\frac{5H^2K}{4}-\frac{H^3K}{4}+\frac{21K^2}{8}
+\frac{21HK^2}{8}\nn\\
&\quad
+\frac{3H^2K^2}{4}-\frac{11K^3}{8}
-\frac{3HK^3}{4}+\frac{K^4}{4} ,\nn\\
I_4(\xi)
&=
-\frac{11}{8}-\frac{17H}{8}-\frac{H^2}{2}
+\frac{17K}{4}+5HK+\frac{3H^2K}{2}+\frac{H^3K}{4}
-\frac{9K^2}{2}-\frac{29HK^2}{8}\nn\\
&\quad
-H^2K^2+\frac{7K^3}{4}
+\frac{3HK^3}{4}-\frac{K^4}{8},\nn\\
I_5(\xi)
&=
-1-\frac{3H}{2}-\frac{3H^2}{4}+\frac{13K}{4}
+\frac{31HK}{8}+\frac{3H^2K}{2}+\frac{H^3K}{4}
-\frac{15K^2}{4}-\frac{13HK^2}{4}\nn\\
&\quad
-\frac{3H^2K^2}{4}+\frac{7K^3}{4}+\frac{7HK^3}{8}-\frac{K^4}{4},
\nn\\
I_6(\xi)
&=
2-H+\frac{3H^2}{4}-\frac{19K}{8}-\frac{3HK}{4}
-\frac{3\,H^2\,K}{8}-\frac{H^3K}{8}-K^2+\frac{HK^2}{2}\nn\\
&\quad
+\frac{H^2K^2}{4}+\frac{5K^3}{4}+\frac{5HK^3}{4}
+\frac{3H^2K^3}{8}+\frac{K^5}{8},\nn\\
I_7(\xi)
&=
\frac{7}{4}+\frac{7H}{4}-\frac{43K}{8}-\frac{39HK}{8}
-\frac{11H^2K}{8}-\frac{H^3K}{8}+\frac{23K^2}{4}
+\frac{19HK^2}{4}\nn\\
&\quad
+\frac{7H^2K^2}{4}+\frac{H^3K^2}{4}
-\frac{5K^3}{2}-\frac{15HK^3}{8}-\frac{3H^2K^3}{8}
+\frac{K^4}{2}+\frac{HK^4}{4}-\frac{K^5}{8}.
\label{inhomterm}
\end{align}

We can solve the differential equations (\ref{diffeq}) by the same
polynomial assumption as used in the construction of the
noncommutative $1/4$ BPS solution \cite{HM}, that is,
we assume that the functions $\calF=\phi_k$ and $a_k$ are given as
polynomials of $H$ and $K$:
\begin{equation}
\calF=\sum_{n=0}^{n_{{\rm max}}}\sum_{m=0}^{m_{{\rm max}}}
\calF_{nm} H^nK^m,
\label{polansatz}
\end{equation}
with suitably large $n_{\rm max}$ and $m_{\rm max}$.
This assumption is owing to the property of $H$ and $K$,
\begin{align}
\calD K &=-HK ,\nn\\
\calD H &= 1 + H - K^2 ,
\end{align}
which implies that the operation of ${\cal D}$ on a polynomial of $H$
and $K$ just reproduces another polynomial of them.
With the assumption (\ref{polansatz}), the differential equations
(\ref{diffeq}) are reduced to a set of linear algebraic equations
for the coefficients $\calF_{nm}$, which can be solved
straightforwardly. Note that we originally had seven differential
equations (\ref{diffeq}) for eight unknown functions, so the solution
contains one undetermined function.
This is the gauge freedom which preserves the generalized rotational
invariant form (\ref{t2 sol}), and the corresponding gauge
transformation function is
\begin{equation}
\lambda^a =\epsilon_{aij}\theta_i{}x_j\tx\frac{1}{r^4}\lambda(\xi).
\end{equation}
Using this freedom to choose $a_5(\xi)=0$, the solution to the
the $\calO(\theta^2)$ part (\ref{SU2}) of the noncommutative BPS
equation are given as follows:
\begin{align}
\phi_1(\xi)&=
-\frac{1}{4}H
+\frac{1}{4}H^{2}
-\frac{1}{8}H^{3}
+\frac{1}{4}HK^{2},\nn\\
\phi_2(\xi)&=
\frac{1}{8}
-\frac{3}{8}H
+\frac{1}{8}H^{2}
-\frac{1}{4}K^{2}
+\frac{3}{8}HK^{2}
+\frac{1}{8}H^{2}K^{2}
+\frac{1}{8}K^{4},\nn\\
\phi_3(\xi)&=
-\frac{1}{8}
+\frac{7}{8}H
-\frac{5}{8}H^{2}
+\frac{1}{8}H^{3}
+\frac{1}{4}K^{2}
-\frac78 HK^{2}
-\frac{1}{8}H^{2}K^{2}
-\frac{1}{8}K^{4}. \nn\\
a_1(\xi)&=
-\frac{1}{8}
+\frac{1}{2}H
-\frac{1}{8}K
-\frac{1}{2}HK
-\frac{1}{4}H^{2}K
+\frac{5}{8}K^{2}
+\frac{1}{4}HK^{2}
-\frac{3}{8}K^{3}
-\frac{1}{4}HK^{3},\nn\\
a_2(\xi)&=
\frac{1}{4}
+\frac{1}{2}H
-\frac{3}{8}H^{2}
-\frac{3}{4}K
-\frac{1}{2}HK
+\frac{1}{8}H^{2}K
+\frac{3}{4}K^{2}
-\frac{1}{4}K^{3},\nn\\
a_3(\xi)&=
-\frac{1}{8}
-\frac{1}{4}H
-\frac{1}{8}H^{2}
+\frac{3}{8}K
+\frac{1}{2}HK
+\frac{1}{8}H^{2}K
-\frac{3}{8}K^{2}
-\frac{1}{4}HK^{2}
+\frac{1}{8}K^{3},\nn\\
a_4(\xi)&=
-\frac{1}{4}
-\frac{3}{2}H
+\frac{1}{2}H^{2}
+\frac{5}{4}K
+\frac{3}{2}HK
+\frac{1}{4}H^{2}K
-\frac{7}{4}K^{2}
-\frac{1}{4}HK^{2}
+\frac{3}{4}K^{3}
+\frac{1}{4}HK^{3}.
\end{align}

\section{SW map and the eigenvalues of the scalar field}
Having obtained the classical solution to the noncommutative BPS
equation to ${\cal O}(\theta^2)$, our next task is to transform it
into the commutative description via the SW map to get the
eigenvalues of the scalar field \cite{HaHi}.
For this purpose, we first have to establish the SW map to second
order in the change $\delta\theta$ of the noncommutativity parameter.

It was pointed out in \cite{AK} that the SW map has inherent
ambiguities. There are two types of ambiguities in it.
One is of the form identifiable as gauge transformations.
The other type of ambiguity consists of gauge covariant quantities.
The latter can cause physical differences and must be fixed by some
physical requirements.
This type of ambiguities comes from the path dependence of the
map in the $\theta$-space.
In other words, even if we perform the map to go round in the
$\theta$-space, we do not come back to the original configuration.
This means that the SW map at $\calO(\delta\theta^2)$ and
higher has such a type of ambiguities.

The SW map for the gauge field to $\calO(\delta\theta^2)$,
including the ambiguities, is presented in Appendix A.
Here, we need the SW map for the scalar field from a
noncommutative space with small $\theta$ to the commutative space.
This map is obtained by performing the dimensional reduction of the
map for the gauge field and taking
\begin{equation}
\delta\theta^{ij}=-\theta^{ij}.
\end{equation}
Then, the scalar field $\Phi$ in the commutative description is
expressed in terms of $\hPhi$ and $\hA_i$ in the noncommutative
description as
\begin{equation}
\Phi = \hPhi + \Delta\hPhi^{(1)} + \Delta\hPhi^{(2)},
\label{Phi=hPhi+DeltaPhi}
\end{equation}
with $\Delta\hPhi^{(1)}$ and $\Delta\hPhi^{(2)}$ given by
\begin{align}
\Delta\hPhi^{(1)}
&=
-\frac14\delta\theta^{kl}\PB{A_k}{(\p_l+D_l)\Phi}
-i\alpha\delta\theta^{kl}\CO{\Phi}{F_{kl}}
-2\beta\delta\theta^{kl}\CO{\Phi}{A_kA_l}
,\label{t1}\\
\Delta\hPhi^{(2)}
&=
\frac{1}{4}\I(\wick{21}{<1\p <2A\,>2\p >1\p\Phi})
+\frac{1}{4}\R(\wick{11}{<1A\,>1\p <2A\, >2\p\Phi})
-\frac{1}{4}\R(\wick{12}{<1\p <2A\, >1A\, >2\p\Phi})
\nn\\
&\quad
-\frac{1}{4}\R(\wick{11}{<1A\,>1\p <2D\Phi\, >2A})
+\frac{1}{4}\R(\wick{12}{<1A\, <2A\, >1\p >2D\Phi})
+\frac{1}{4}\R(\contr{AF}{4}{5}{4}\contr{D}{-1}{3}{4}\Phi)
\nn\\
&\quad
+\frac{1}{4}\R(\contr{A}{4}{-30}{8}\,\contr{D}{5}{-14}{4}\Phi\,F)
-\frac{1}{8}\R(\wick{21}{<1\p <2A\, >2\p >1A\,\Phi})
+\frac{1}{8}\R(\wick{21}{<1\p <2A\,\Phi\, >2\p >1A})
\nn\\
&\quad
-\frac{1}{8}\R(\wick{12}{<1A\, <2A\, >1A\, >2A\,\Phi})
+\frac{1}{4}\R(\wick{12}{<1A\, <2A\, >1A\,\Phi\, >2A})
-\frac{1}{8}\R(\wick{12}{<1A\, <2A\,\Phi\, >1A\, >2A})
\nn\\
&\quad
-\Bigl(\frac{1}{16}+8\alpha\beta+4\beta^2\Bigr)
\R(\wick{11}{<1A>1A\,[\Phi,<2A>2A]})
+8\alpha\beta
\I(\wick{11}{<1\p>1A\,[\Phi,<2A>2A]})
\nn\\
&\quad
-\gamma_1\R(\contr{F}{3}{-27}{8}\,[\Phi,\contr{F}{-21}{5}{4}])
-\gamma_2\R(\scF\,[\Phi,\scF])
+\gamma_3\I(\wick{1}{<1D\Phi\,>1D}\!\scF)
\nn\\
&\quad
+\Delta\hPhi_{{\rm metric}}^{(2)}
+(\mbox{gauge-type ambiguities}) .
\label{t2}
\end{align}
In (\ref{t1}) and (\ref{t2}), all the fields are defined at $\theta$
and all the products are the $*$ products (we have omitted the hats on
the fields and the $*$ for the products).
We have used the following simplified notation:
\begin{equation}
\wick{1}{<1A>1B}\equiv\delta\theta^{kl}A_kB_l,\quad
\wick{1}{<1 A >1 F_i}\equiv \delta\theta^{kl}A_k F_{li},\quad
\scF\equiv\delta\theta^{kl}F_{kl},
\label{contraction}
\end{equation}
and
\begin{equation}
\R\calO \equiv \frac12\left(\calO + \calO^\dagger\right),
\quad
\I\calO \equiv \frac{1}{2i}\left(\calO - \calO^\dagger\right).
\label{ReIm}
\end{equation}
Note that the contraction symbol has the property that
$(\wick{1}{<1A>1B})^\dagger = -\wick{1}{<1B^\dagger>1A^\dagger}$
due to the anti-symmetry of $\delta\theta^{kl}$.
We shall mention the $\Delta\hPhi^{(2)}_{\rm metric}$ term in
(\ref{t2}) soon below.

The SW map at $\calO(\delta\theta)$ (\ref{t1}) contains
two ambiguities parameterized by $\alpha$ and $\beta$.
They are the gauge-type ambiguities.
On the other hand, there exist three covariant-type ambiguities
with coefficients $\gamma_k$ ($k=1,2,3$) in the SW map at
$\calO(\delta\theta^2)$ (\ref{t2}).
This type of ambiguities directly affects the eigenvalues of the
scalar.
Note that the gauge-type ambiguities at $\calO(\delta\theta)$ have
influence on the map at $\calO(\delta\theta^2)$ and may possibly
change the eigenvalues.

The term $\Delta\hPhi^{(2)}_{\rm metric}$ in (\ref{t2}) represents the
covariant-type ambiguities using the metric.\footnote{
SW map containing the metric was also considered in \cite{OT} in a
different context.
}
Note that, in all the terms written explicitly in (\ref{t2}), the
upper indices of $\theta^{ij}$ are contracted with the lower ones of
$A_i$ and $\p_i$ without using the metric.
On the other hand, the terms in $\Delta\hPhi^{(2)}_{\rm metric}$ are
constructed by using the metric.
There are many terms belonging to $\Delta\hPhi^{(2)}_{\rm metric}$.
Examples are
\begin{equation}
\delta^{km}\delta^{ln}\R(F_{kl}\,\CO{\Phi}{F_{mn}})\,\QQ,
\quad
\delta^{km}\delta^{ln}
\R(D_k\Phi\,\CO{\Phi}{D_l\Phi})\,\theta_m\theta_n.
\label{Phimetric}
\end{equation}
These are obtained by the dimensional reduction of the corresponding
operators in the SW map for the gauge field given in eq.\
(\ref{Ametric}) of Appendix A.

Now we shall proceed to the evaluation of the eigenvalues of the
scalar $\Phi$ (\ref{Phi=hPhi+DeltaPhi}) as a $2\times 2$ matrix
to $\calO(\theta^2)$.
For this purpose, we expand $\Phi$ in the commutative description in
powers of $\theta$:
\begin{equation}
\Phi=\Phi^{(0)} + \Phi^{(1)} +\Phi^{(2)} ,
\label{P=P0+P1+P2}
\end{equation}
where $\Phi^{(n)}$ is of order $\theta^n$.
Let us write explicitly the arguments of the SW map, (\ref{t1}) and
(\ref{t2}), as $\Delta\hPhi^{(n)}[\hPhi,\hA_i,\theta]$ with the last
argument representing the $\theta$-dependence only through
the $*$ product (\ref{star}). Then, using the noncommutative classical
solution (\ref{expand}), we have
\begin{align}
\Phi^{(0)} &=\hPhi^{a(0)}\frac12\sigma_a ,
\label{P0}
\\
\Phi^{(1)} &=\Delta\hPhi^{(1)}[\hPhi^{(0)},\hA_i^{(0)},\theta=0] ,
\label{P1}
\\
\Phi^{(2)} &=\hPhi^{a(2)}\frac12\sigma_a +
\Delta\hPhi^{(1)}[\hPhi,\hA_i,\theta]\Big|_{\theta^2}
+\Delta\hPhi^{(2)}[\hPhi^{(0)},\hA_i^{(0)},\theta=0] .
\label{P2}
\end{align}
We shall add some explanations about (\ref{P0}) -- (\ref{P2}).
First, we have set $\theta=0$ in (\ref{P1}) and the last term of
(\ref{P2}).
This implies that we take the commutative products among the
fields.
Next, the second term on the RHS of (\ref{P2}) means the sum of all
the terms quadratic in $\theta$ in
$\Delta\hPhi^{(1)}[\hPhi,\hA_i,\theta]$.
There are three sources of $\theta$;
$\delta\theta^{kl}=-\theta^{kl}$ in (\ref{t1}), $\theta$ in the
$*$ product, and $\theta$ in the noncommutative classical solution
(\ref{t1 sol}).

Then, the two eigenvalues $\lambda_\pm$ of the scalar $\Phi$
(\ref{P=P0+P1+P2}) are given using the well-known perturbation theory
formula as
\begin{equation}
\lambda_\pm=\lambda_\pm^{(0)}+\lambda_\pm^{(1)}+ \lambda_\pm^{(2)} ,
\end{equation}
with
\begin{align}
\lambda^{(0)}_\pm &= \pm\frac{H(\xi)}{2r} ,
\\
\lambda^{(1)}_\pm &=\bra{\pm}\Phi^{(1)}\ket{\pm} ,
\label{eigenvalue1}
\\
\lambda_\pm^{(2)} &=
\bra{\pm}\Phi^{(2)}\ket{\pm}
+\frac{\bra{\pm}\Phi^{(1)}\ket{\mp}
\bra{\mp}\Phi^{(1)}\ket{\pm}}{\lambda_\pm^{(0)}-\lambda_\mp^{(0)}},
\label{eigenvalue2}
\end{align}
where the kets $\ket{\pm}$ are the eigenvectors of $\Phi^{(0)}$
satisfying
\begin{equation}
\hx\cdot\sigma\ket{\pm}=\pm\ket{\pm} .
\end{equation}

Plugging the noncommutative classical solution obtained in section 2
into (\ref{eigenvalue1}) and (\ref{eigenvalue2}), we get after a
tedious but straightforward calculation
\begin{align}
\lambda^{(1)}_+
&=-\frac{1}{4r^3}H(1-K^2)\tx ,
\label{l1}
\\[1.5ex]
\lambda^{(2)}_+
&=\frac{1}{16 r^5}\left(
H^2- HK^2+\frac{1}{2}H^3K^2+HK^4\right)\QQ\nn\\
&\quad
+\frac{1}{16 r^5}
\left(
2 H- 3 H^2- HK^2 -\frac{1}{2}H^3K^2-HK^4\right)\tx^2
\nn\\
&\quad
+ c_1\,f_1(x,\theta) +c_2\,f_2(x,\theta)
+\bra{+}
\Delta\hPhi^{(2)}[\hPhi^{(0)},\hA_i^{(0)},\theta=0]_{\rm metric}
\ket{+},
\label{l2}
\end{align}
with
\begin{equation}
c_1=-\frac12\gamma_1 +\gamma_2 -\frac12\gamma_3 +\alpha^2,
\qquad
c_2=\frac12\gamma_3 ,
\label{c12}
\end{equation}
and
\begin{align}
f_1(x,\theta) &=\frac{1}{r^5}H^3K^2\left(\QQ-\tx^2\right) ,
\nn\\
f_2(x,\theta) &=\frac{1}{r^5}
\left(HK^2-H^2K^2-HK^4\right)\left(\QQ-3\tx^2\right) .
\label{f12}
\end{align}
The other eigenvalue $\lambda_-$ is given by $\lambda^{(1)}_-=
\lambda^{(1)}_+$ and $\lambda^{(2)}_-=-\lambda^{(2)}_+$.
The first order eigenvalue (\ref{l1}) is already obtained in
\cite{HaHi}.
The origin of the $\alpha^2$ term in $c_1$ (\ref{c12}) is the last
term of (\ref{eigenvalue2}).
There are no other contributions to $\lambda^{(2)}$ from the last term
of (\ref{eigenvalue2}) since
we have $\Phi^{(1)}|_{\alpha=\beta=0}\propto\unit$.
The terms in the SW map (\ref{t2}) quadratic in $\alpha$ and $\beta$
do not contribute to $\lambda^{(2)}$ owing to the property
$[\Phi^{(0)},\wick{1}{<1A^{(0)}>1A^{(0)}}]=0$ for the zero-th order
solution.
All the constituents of $\lambda^{(2)}_+$ (\ref{l2}), which are
polynomials of $H$ and $K$ divided by $r^5$, vanish at the origin
$r=0$.

\section{Tilted D-string picture}

We would like to compare the the scalar eigenvalues obtained in the
previous section with those which are obtained by
different ways and are expected to describe the same physical
situation of the tilted D-string between two parallel D3-branes.
In the $U(1)$ case, there are three ways giving the same result
\cite{HaHi,Mori};
the SW map of the noncommutative BPS solution,
the nonlinear BPS solution in the commutative space,
and the target space rotation of the linear BPS solution in the
commutative space.
In particular, the linear BPS solution (under a constant magnetic
field) gives the tilted D3-brane picture, which is related to the
tilted D-string picture by the target space rotation (see figure 1).
In the nonabelian case, the nonlinearly realized
supertransformation of the DBI theory is not
well-understood. Therefore, we shall take the target space rotation of
the linear BPS solution in the commutative space as the object to be
compared with the eigenvalues of section 3.

\begin{figure}[hbt]
\begin{center}
\leavevmode
\epsfxsize=120mm
\put(117,90){ROTATION}
\put(62,78){D-string}
\put(80,10){D3-brane}
\put(310,70){D-string}
\put(300,0){D3-brane}
\epsfbox{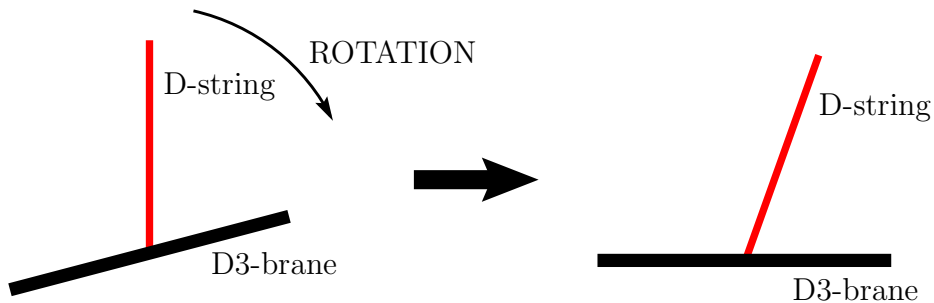}
\caption{When the gauge group is $U(1)$, the target space rotation
  precisely connects the tilted D3-brane picture (left) with the tilted
  D-string picture (right).}
\label{u1mp}
\end{center}
\end{figure}

Let us consider the $U(2)$ super Yang-Mills theory in the
commutative space with a constant $U(1)$ magnetic field $B_i$.
The BPS equation of this system, which we regard as describing the
tilted D3-brane in a constant NSNS $B$-field
$B_{ij}=\epsilon_{ijk}B_k$,
is
\begin{equation}
D_i\Phi+\frac{1}{2}\epsilon_{ijk}\left(
F_{jk}+B_{jk}\frac{1}{2}\unit\right)=0.
\label{commBPS}
\end{equation}
For our present purpose of comparing with the previous section, we
should in fact consider the Yang-Mills theory in the commutative space
with the metric $g_{ij}$ related to the metric $G_{ij}=\delta_{ij}$ in
the noncommutative theory of section 2 by \cite{SW}
\begin{equation}
G^{ij}+\frac{\theta^{ij}}{2\pi\alpha'}
=\left(\frac{1}{g+2\pi\alpha' B}\right)^{ij} .
\end{equation}
However, the desired scalar eigenvalue is obtained by considering the
BPS equation (\ref{commBPS}) with $g_{ij}=\delta_{ij}$ and
coordinate transforming back to the original $g_{ij}$
afterwards \cite{Mori}.

The $U(1)$ part of this equation
\begin{equation}
\p_i\Phi^0+\frac{1}{2}\epsilon_{ijk}B_{jk}=0,
\end{equation}
is easily solved to give
\begin{equation}
\Phi^0
=-\frac{1}{2}\epsilon_{ijk}B_{jk}x^i
=\frac{1}{(2\pi\alpha')^2}(\theta{}x) ,
\end{equation}
where the relation
$2\pi\alpha'B_i=-\theta_i/2\pi\alpha'+{\cal O}(\theta^3)$ has been
used.
As a solution to the nonabelian part, we adopt the ordinary BPS
monopole solution (\ref{t0 sol}).
We shall attach tilde to the space coordinates in the present
system for distinguishing them from those in the rotated system to
be discussed below.
Then, the (larger) eigenvalue of the scalar field is
\begin{equation}
\wt{\ev}=
\frac{1}{2\wt{r}}H(\wt{\xi})
+\frac{1}{(2\pi\alpha')^2}(\theta\wt{x}) ,
\end{equation}
with $\wt{\xi}\equiv\wt{C}\wt{r}$ and $\wt{r}\equiv\wt{x}_i\wt{x}_i$
($\wt{C}$ is the mass scale of the present monopole).

\begin{figure}[hbt]
\begin{center}
\leavevmode
\epsfxsize=170mm
\put(195,102){$(\wh{\theta}\wt{x})$}
\put(180,148){$(\wh{\theta}x)$}
\put(96,178){$\wt{\Lambda}$}
\put(59,173){$\Lambda$}
\put(40,8){D3-brane}
\put(-10,115){D3-brane}
\put(95,149){$\wt{C}$}
\put(110,35){$-\wt{C}$}
\put(74,137){$C$}
\put(133,48){$-C$}
\put(98,-17){{\Large (A)}}
\put(144,102){$\phi$}
\put(470,102){$(\wh{\theta}x)$}
\put(377,185){$\Lambda$}
\put(300,145){D3-brane}
\put(300,38){D3-brane}
\put(378,145){$C$}
\put(370,38){$-C$}
\put(380,-17){{\Large (B)}}
\epsfbox{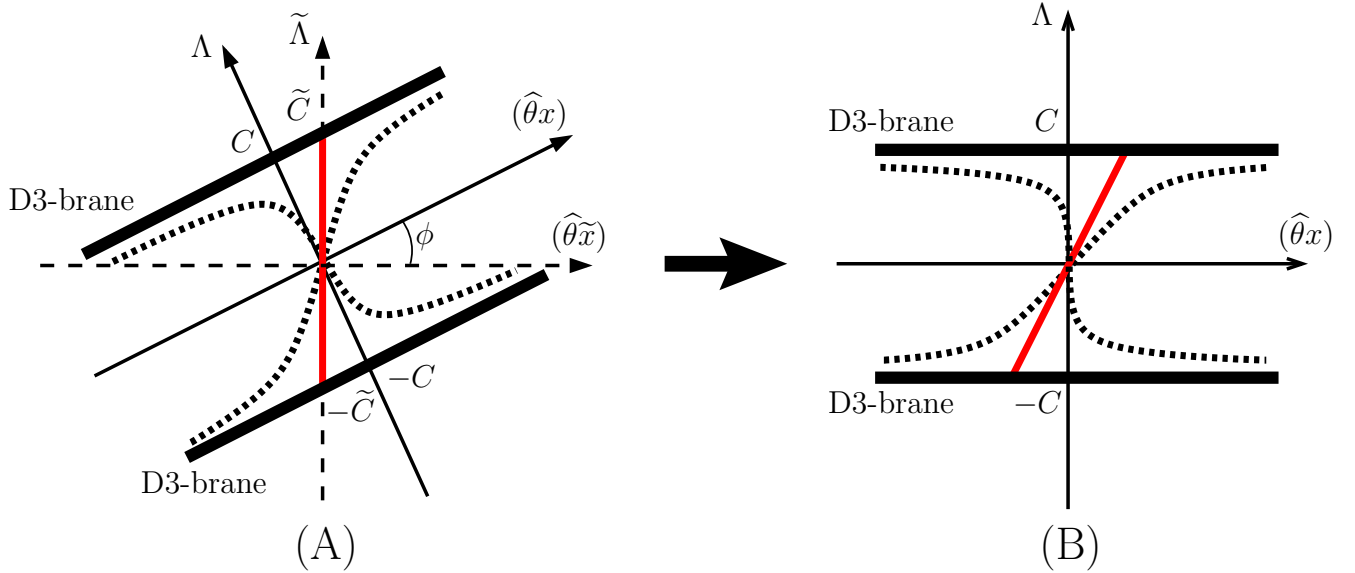}
\caption{The tilted D3-brane picture (A) in the $U(2)$ case is expected
  to be related to the tilted D-string picture (B) by a target space
  rotation. The dotted curves represent the scalar eigenvalues.
We have omitted $2\pi\alpha'$ which should multiply $\Lambda$,
  $\wt{\Lambda}$, $C$ and $\wt{C}$ in the two figures.
}
\label{u2mp}
\end{center}
\end{figure}

Now let us carry out the target space rotation and turn to the
tilted D-string picture (see figure \ref{u2mp}):
\begin{equation}
\left(
\begin{array}{@{\,}c@{\,}}
2\pi\alpha' \ev\\
(\wh{\theta}x)
\end{array}
\right)
=\left(
\begin{array}{@{\,}cc@{\,}}
\cos\phi & -\sin\phi\\
\sin\phi & \cos\phi
\end{array}
\right)
\left(
\begin{array}{@{\,}c@{\,}}
2\pi\alpha' \wt{\ev}\\
(\wh{\theta}\wt{x})
\end{array}
\right) ,
\end{equation}
where $\wh{\theta}_i$ is the unit vector
$\wh{\theta}_i\equiv\theta_i/|\theta|$, and the rotation angle $\phi$
is given as  $\tan\phi=|\theta|/2\pi\alpha'$.
The components perpendicular to $\theta_i$ are common between $x_i$
and $\wt{x}_i$.
Expressing the new eigenvalue $\ev$ in the tilted D-string picture
as a function of the coordinate $x^i$, we have
\begin{align}
\ev
&=
\frac{H}{2r}
-\frac{1}{4r^3}H(1-K^2)\tx
+\frac{1}{16r^5}\left(H^2-H^2K^2\right)\QQ
\nn\\
&\quad
+\frac{1}{16r^5}\left(2H-3H^2-4HK^2+3H^2K^2+2H^3K^2
+2HK^4\right)\tx^2
\nn\\
&\quad
-\frac{1}{(2\pi\alpha')^2}\frac{1}{4r}\left\{
H\QQ +(1-K^2)\tx^2\right\} ,
\label{nl mono}
\end{align}
where the arguments of $H$ and $K$ are $\wt{C}r$ with
$r^2=x^ix^i$.
Now we make the coordinate transformation in (\ref{nl mono})
from the metric $\delta_{ij}$ to
$g_{ij}=\delta_{ij}-(\QQ\delta_{ij}-\theta_i\theta_j)/(2\pi\alpha')^2$
corresponding to
the open string metric $G_{ij}=\delta_{ij}$ adopted in section 2.
This is accomplished by the replacement of $r$ with
\begin{equation}
\left(g_{ij}x^ix^j\right)^{1/2}=\left(
1-\frac{1}{(2\pi\alpha')^2}\frac{1}{2}
\left[\QQ-\tx^2\right]\right)r .
\label{R}
\end{equation}
Then, the eigenvalue $\Lambda$ in the new coordinate system is given
by (\ref{nl mono}) with the last $1/(2\pi\alpha')^2$ term omitted and
the arguments of $H$ and $K$ replaced by $Cr$ (cf.\ \cite{Mori} for
the $U(1)$ case).
Here, $C$ is the the D3-brane separation, $C\equiv\wt{C}\cos\phi$
(see figure \ref{u2mp}).
We can show that this eigenvalue is exactly the same as $\ev(x)$
obtained by solving
\begin{equation}
\ev(x)=\frac{1}{2r}H(|x_i-\ev(x)\theta_i|) ,
\label{L=HL}
\end{equation}
which implies
the tilted D-string picture in figure \ref{u2mp}(B).
Namely, for a given value of $\Lambda$, the corresponding $x_i$ lies on
a sphere with its center at $x_i=\Lambda\theta_i$
(cf.\ \cite{HHM,HM,HaHi}).

Having finished the preparation of obtaining the eigenvalue from the
target space rotation of the linear BPS solution, let us proceed to
the comparison between this eigenvalue $\Lambda$ (\ref{nl mono})
(without the $1/(2\pi\alpha')^2$ term) and the eigenvalue (\ref{l1})
and (\ref{l2}) obtained from the noncommutative monopole via the SW
map.
First, the $\calO(\theta)$ terms agree between them as was already
shown in \cite{HaHi}.
Second, the $\calO(\theta^2)$ parts coincide perfectly in
the asymptotic region $r\to\infty$ where we can drop the exponentially
decaying terms (see eq.\ (\ref{asympt})).
Note in particular that all the ambiguity terms in (\ref{l2})
disappear in the the asymptotic region.
(The last term of (\ref{l2}) using the metric is also exponentially
decaying as $r\to\infty$.)

Let us compare the $\calO(\theta^2)$ terms in the two
eigenvalues for a general $x_i$ not restricted to the asymptotic
region.
Since the SW map is defined by the gauge equivalence relation
independent of the metric, we shall consider first the simpler case of
(\ref{l2}) without the last
term $\bra{+}\Delta\hPhi^{(2)}_{\rm metric}\ket{+}$ using the metric.
In this case, by taking $c_2=1/16$, we can make (\ref{l2}) agree with
the $\calO(\theta^2)$ part of (\ref{nl mono}) except only the
$H^3K^2$ terms.
However, for the complete agreement between the two eigenvalues, the
introduction of the metric term
$\bra{+}\Delta\hPhi^{(2)}_{\rm metric}\ket{+}$ is inevitable.

As we mentioned in section 3, there are many contributions to the
covariant-type ambiguity using the metric. A complete
analysis shows that the term
$\bra{+}\Delta\hPhi^{(2)}_{\rm metric}\ket{+}$
is a sum of three functions, $f_1$ and $f_2$ of (\ref{f12}) and a new one
\begin{equation}
f_3(x,\theta)=\frac{1}{r^5}H^3K^2\QQ ,
\label{f3}
\end{equation}
each multiplied by an arbitrary coefficient.
In fact, we have
$\bra{+}\delta^{km}\delta^{ln}\R(F_{kl}\,\CO{\Phi}{F_{mn}})\,\QQ\ket{+}
= - f_3(x,\theta)$.
Then, expressing the RHS of $\lambda^{(2)}_+$ (\ref{l2}) as the
sum of its first two terms and $c_1f_1+c_2f_2+c_3f_3$ with the
redefined $c_1$ and $c_2$, the complete
agreement between $\lambda^{(2)}_+$ and the $\calO(\theta^2)$ term of
$\Lambda$ (\ref{nl mono}) is achieved by taking the three parameters as
$c_1=-5/32$, $c_2=1/16$ and $c_3=1/8$.
This is the unique choice for the coefficients $c_k$.
Note that this agreement is a non-trivial one since we have to tune
eight coefficients by using only three free parameters.
Of course, the three coefficients $c_k$ do not completely fix
the ambiguity in the SW map since there are many contributions to
$c_k$ if we allow the $\Delta\hPhi^{(2)}_{\rm metric}$ term using the
metric. The use of the metric in the SW map seems not so unnatural if
we recall that the noncommutative classical solution
(cf.\ (\ref{t1 sol}) and (\ref{t2 sol})) as well as the BPS equation
(\ref{BPSeq}) already contains the metric.

\section{Summary and discussions}

In this paper, we considered the noncommutative monopole solutions at
the second order in $\theta$.
We solved the noncommutative version of the BPS equation to
$\calO(\theta^2)$, mapped the solution to the commutative side, and
obtained the eigenvalues of the resulting scalar field.
We saw that the ambiguities in the SW map have explicit influence
on the scalar eigenvalues.
We made the brane interpretation to the scalar eigenvalues and
examined whether they can reproduce the configuration of a tilted
D-string suspended between two parallel D3-branes.
In the asymptotic region, the effect of the ambiguities in the SW map
disappear and at the same time the scalar eigenvalue precisely give the
expected D-string picture.
Without the restriction to the asymptotic region, we found that we can
tune the free parameters in the SW map so that the scalar eigenvalues
reproduce the desired configuration.
It is necessary to introduce the covariant-type ambiguity term
using the metric.
The number of free parameters $c_k$ in the eigenvalues is just enough
to adjust them to the expected ones.

We would like to make a few comments.
Our first comment is on the covariant type ambiguity in the SW map.
In this paper we have constructed the SW map first in the pure
Yang-Mills system without the scalar field and then obtained the map
for the scalar by the dimensional reduction of the map for the gauge
field. This is natural if we recall the origin of the present super
Yang-Mills  theory via the dimensional reduction.
However, if we forget this origin, there are other covariant-type
ambiguities treating the scalar field $\Phi$ as a gauge covariant
quantity from the start. For example, as an ambiguity for the scalar
field at $\calO(\theta)$, we have
$\delta\theta^{ij}\PB{\Phi}{F_{ij}}$.
However, this term gives the same contribution (with an arbitrary
coefficient) to the $\calO(\theta)$ eigenvalue as the existing one
(\ref{l1}), and hence even the tilt angle at $\calO(\theta)$ becomes a
free parameter.

Next we shall comment on the noncommutative eigenvalue equation for
the scalar field proposed and examined in \cite{HHM,HM}.
At $\calO(\theta)$, the eigenvalues of the noncommutative eigenvalue
problem for the scalar gave the same asymptotic behavior as those
obtained via the SW map \cite{HaHi}. We have carried out the analysis
of the noncommutative eigenvalue equation for the classical solution
at $\calO(\theta^2)$ given in section 2.
However, the resulting eigenvalues do not agree with those from the SW
map even in the asymptotic region.
Therefore, the noncommutative eigenvalue equation seems to work well
only at the first order in $\theta$, though it is still an interesting
subject to understand why it gives a good result at this order.

Finally, we would like to emphasize the usefulness of the analysis
using the BPS solutions. The BPS solutions are expected to remain
intact even if we include the $\alpha'$ corrections.
Thus, the BPS solutions would be helpful for giving
a support for the equivalence between the noncommutative description
and the commutative one independently of the $\alpha'$ expansion.
It is a very interesting subject to pursue the method which enables us
to examine this equivalence to all orders in $\theta$.

\section*{Acknowledgments}
We would like to thank T.\ Asakawa, K.\ Hashimoto, I.\ Kishimoto and
S.\ Moriyama for valuable discussions and useful comments.
This work is supported in part by Grant-in-Aid for Scientific Research
from Ministry of Education, Science, Sports and Culture of Japan
(\#03602 and \#12640264).
The work of S.\ G.\ is supported in part by the Japan Society for the
Promotion of Science under the Predoctoral Research Program.

\appendix
\section{Seiberg-Witten map to $\calO(\delta\theta^2)$}

In this appendix, we present the SW map for the gauge field to
second order in the change $\delta\theta$ of the noncommutativity
parameter $\theta$.
The SW map is derived from the gauge equivalence relation \cite{SW}
\begin{equation}
A_i(\hA)+\delta_{\lambda}A_i(\hA)
=A_i(\hA+\wh{\delta}_{\hl}\hA),
\label{gauge equiv}
\end{equation}
where the quantities with a hat are defined at $\theta$ and those
without hat at $\theta+\delta\theta$, a nearby point of $\theta$.
This unconventional meaning of hat is for the convenience of the use
in section 3.

We expand $A_i$ and $\lambda$ in powers of $\delta\theta$:
\begin{align}
A_i &=\hA_i+\Delta\hA_i^{(1)}+\Delta\hA_i^{(2)}
+\calO(\delta\theta^3),\nn\\
\lambda &=\hl+\Delta\hl^{(1)}+\Delta\hl^{(2)}
+\calO(\delta\theta^3).
\end{align}
Substituting them into (\ref{gauge equiv}), the first order part is
solved in the most general form as \cite{AK}
\begin{align}
\Delta\hA_i^{(1)}
&=
-\frac{1}{4}\delta\theta^{kl}\PB{\hA_k}{\p_l\hA_i+\hF_{li}}
+\alpha\delta\theta^{kl}\hD_i\hF_{kl}
-i\beta\delta\theta^{kl}\hD_i[\hA_k,\hA_l] ,\\
\Delta\hl^{(1)}
&=
-\frac{1}{4}\delta\theta^{kl}\PB{\hA_k}{\p_l\hl}
-2i\beta\delta\theta^{kl}\CO{\hA_k}{\p_l\hl} ,
\end{align}
where $\alpha$ and $\beta$ are arbitrary real coefficients.
Note that these two ambiguity terms are both gauge-type ones.
Next, we shall solve the second order part of the equation
(\ref{gauge equiv}),
\begin{align}
&\wh{\delta}_{\hl}\Delta\hA_i^{(2)}
+i[\Delta\hA_i^{(2)},\hl]
-\hD_i\Delta\hl^{(2)}
=
\frac{i}{8}\delta\theta^{kl}\delta\theta^{mn}
[\p_k\p_m\hA_i,\p_l\p_n\hl]\nn\\
&\qquad
+\frac{1}{2}\delta\theta^{kl}\PB{\p_k{}\hA}{\p_l\Delta\hl^{(1)}}
+\frac{1}{2}\delta\theta^{kl}\PB{\p_k\Delta\hA_i^{(1)}}{\p_l\hl}
-i\CO{\Delta\hA_i^{(1)}}{\Delta\hl^{(1)}} ,
\label{t2 for gf}
\end{align}
to obtain $\Delta\hA_i^{(2)}$ and $\Delta\hl^{(2)}$.
We solved this equation (\ref{t2 for gf}) by assuming the most general
forms for $\Delta\hA_i^{(2)}$ and $\Delta\hl^{(2)}$.
The result is as follows:\footnote{
The terms in $\Delta\hPhi^{(2)}$ of the form $\R\calO$ ($\I\calO$)
with $\calO$ containing odd (even) number of derivatives do not
contribute to the scalar eigenvalue formula (\ref{eigenvalue2}).
Therefore, in eq.\ (\ref{DA2}) we have omitted such kind of terms,
which would appear as the covariant-type ambiguity terms and
the terms quadratic in $\alpha$ and $\beta$.
}
\begin{align}
\Delta\hA_i^{(2)}
&=
\frac{1}{4}\I(\wick{21}{<1\p <2A\,>2\p >1\pA_i})
-\frac{1}{8}\I(\wick{21}{<1\p <2A\,>2\p \p_i  >1A})
+\frac{1}{4}\R(\wick{11}{<1A\,>1\p <2A\, >2\pA_i})
-\frac{1}{4}\R(\wick{12}{<1\p <2A\, >1A\, >2\pA_i})
\nn\\
&\quad
-\frac{1}{4}\R(\wick{11}{<1A\,>1\p <2F_i\, >2A})
+\frac{1}{4}\R(\wick{12}{<1A\, <2A\, >1\p >2F_i})
+\frac{1}{4}\R(\contr{AF}{5}{5}{4}\contr{F_i}{-2}{5}{4})
+\frac{1}{4}\R(\contr{A}{5}{-19}{8}\contr{F_i}{5}{-4}{4}F)
\nn\\
&\quad
-\frac{1}{8}\R(\wick{21}{<1\p <2A\, >2\p >1A\,A_i})
+\frac{1}{8}\R(\wick{21}{<1\p <2A\,A_i\, >2\p >1A})
-\frac{1}{8}\I(\wick{12}{<1A\, <2A\, >1A\,\p_i >2A})
+\frac{1}{8}\I(\wick{12}{<1A\, <2A\,\p_i >1A\, >2A})
\nn\\
&\quad
-\frac{1}{8}\R(\wick{12}{<1A\, <2A\, >1A\, >2A\,A_i})
+\frac{1}{4}\R(\wick{12}{<1A\, <2A\, >1A\,A_i\, >2A})
-\frac{1}{8}\R(\wick{12}{<1A\, <2A\,A_i\, >1A\, >2A})
\nn\\
&\quad
+\Bigl(\frac{1}{16}+8\alpha\beta+4\beta^2\Bigr)
\I(\wick{11}{<1 A\, >1A\,D_i(<2 A\, >2A)})
+8\alpha\beta
\R(\wick{11}{<1 \p>1A\,D_i(<2 A\, >2A)})
\nn\\
&\quad
+\gamma_1
\I(\contr{F}{3}{-23}{8}\,D_i\contr{F}{-18}{5}{4})
+\gamma_2
\I(\scF\,D_i \scF)
+\gamma_3
\I(\wick{1}{<1F_i\,>1D}\!\scF)
\nn\\
&\quad
+\Delta\hA_{i\,{\rm metric}}^{(2)}
+(\mbox{gauge-type ambiguities}) ,
\label{DA2}\\[1.5ex]
\Delta\hl^{(2)}
&=
\frac18\I(\wick{21}{<1\p <2A\,>2\p >1\p\lambda})
-\frac18\R(\wick{12}{<1\p <2A\,>1A\,>2\p\lambda})
-\frac18\R(\wick{21}{<1\p <2A\,>2\p\lambda\,>1A})
+\frac14\R(\wick{11}{<1A\,>1\p<2A\,>2\p\lambda})
\nn\\
&\quad
-\frac18\I(\wick{12}{<1A\,<2A\,>1A\,>2\p\lambda})
+\frac18\I(\wick{12}{<1A\,<2A\,>1\p\lambda\,>2A})
\nn\\
&\quad
+\Bigl(\frac{1}{16}+8\alpha\beta+4\beta^2\Bigr)
\I(\wick{1}{<1A >1A}\,(\wick{1}{<1A\,>1\p\lambda}
+\wick{1}{<1\p\lambda\,>1A}))
+8\alpha\beta
\R(\wick{1}{<1 \p >1A}\,(\wick{1}{<1A\,>1\p\lambda}
+\wick{1}{<1\p\lambda\,>1A}))
\nn\\
&\quad
+(\mbox{gauge-type ambiguities}) ,
\label{Dl2}
\end{align}
where the meanings of the contraction, $\R\calO$ and $\I\calO$ are as
given by eqs.\ (\ref{contraction}) and (\ref{ReIm}).
We have omitted hats on the RHS of (\ref{DA2}) and (\ref{Dl2}).
The ambiguities of the SW map at $\calO(\theta^2)$ are the homogeneous
solutions to eq.\ (\ref{t2 for gf}).
The terms in (\ref{DA2}) multiplied by $\gamma_k$ ($k=1,2,3$) are the
covariant-type ambiguities which cannot be identified as gauge
transformation. All other covariant-type terms are reduced to the
three $\gamma_k$ terms owing to the Bianchi identity.
The term $\Delta\hA_{i\,{\rm metric}}^{(2)}$ denotes
the covariant-type ambiguity using the metric $G_{ij}$. There are
many operators belonging to this type; for example,
\begin{equation}
G^{km}G^{ln}\I(F_{kl}D_iF_{mn})\,\theta^2,\quad
G^{kp}G^{mq}G^{ln}\I(F_{kl}D_i F_{mn})\theta_p\theta_q .
\label{Ametric}
\end{equation}
The SW map for the scalar field $\Phi$ used in section 3 is obtained
from (\ref{DA2}) by the dimensional reduction using
$A_{\Phi}=\Phi$, $F_{i \Phi}=D_i\Phi$ and
$D_\Phi\calO=-i[\Phi,\calO]$.
The second quantity in (\ref{Phimetric}) is obtained from that in
(\ref{Ametric}) by setting $i=\Phi$ and taking the $G^{\Phi\Phi}$
part.

\end{document}